# Google Street View image of a house predicts car accident risk of its resident

Kinga Kita-Wojciechowska, Faculty of Economic Sciences, University of Warsaw, Warsaw, Poland

Łukasz Kidziński, Department of Bioengineering, Stanford University, Stanford, CA, USA

**Road traffic injuries are a leading cause of death worldwide. Proper estimation of car accident risk is critical for appropriate allocation of resources in healthcare, insurance, civil engineering, and other industries. We show how images of houses are predictive of car accidents. We analyze 20,000 addresses of insurance company clients, collect a corresponding house image using Google Street View, and annotate house features such as age, type, and condition. We find that this information substantially improves car accident risk prediction compared to the state-of-the-art risk model of the insurance company and could be used for price discrimination. From this perspective, public availability of house images raises legal and social concerns, as they can be a proxy of ethnicity, religion and other sensitive data.**

Modern machine learning techniques for computer vision, such as Deep Learning, provided unprecedented opportunities for academic research and industrial applications. Examples include using satellite images for deforestation monitoring in South America (*1*) or poverty estimation in Africa (*2*), prediction of skin cancer from skin lesion images (*3*), or automatic detection of pulmonary tuberculosis from a chest radiograph (*4*).

One of the resources recently leveraged for research is Google Street View—a platform from Google where images of buildings are taken using cars equipped with a set of cameras (*5*). This data source has recently been explored by researchers to answer questions in social science, e.g. demographic makeup of neighborhoods across the US (*6*), estimating city-level travel patterns in Great Britain (*7*) or crime rate in Brazil (*8*).

Our work explores whether Google Street View images of houses are predictive of the car accident risk of their residents. Proper risk estimation is key for the insurance and healthcare industries. Insurance is a promise to compensate the potential damage or loss in the future for a relatively low price paid now. Unlike commodity products, the ultimate cost of an insurance policy is not known at the time of the sale. It is, therefore, a challenge to set up a proper pricing and insurers try to leverage statistical methods to predict the future risk of each client upfront. For this purpose, insurers collect historical data about underwritten policies and claims incurred and they build statistical models to identify systematic and time-invariant clients' characteristics that correlate with the number of claims. For example, the classical motor insurance risk factors identified worldwide are the age of the driver, the characteristics of his car, the occurrence of car accidents in the past and geography (*9*). This is why the insurers tend to ask for these and other details before providing the motor insurance offer.

Although insurers often collect address information from the client, they typically use only zip-code for risk modeling and pricing purposes. Claims data aggregated to zip-codes are still too volatile and require spatial smoothing (*10*) and further aggregation to larger geographical zones (*11*) . Such commonly used methodology is based on the assumption that neighbors are driving in a similar manner. In this paper, we challenge this assumption and show that volatility can be explained at the granularity of individual addresses. Moreover, we show that this information can be extracted from publicly available images from the Google Street View.

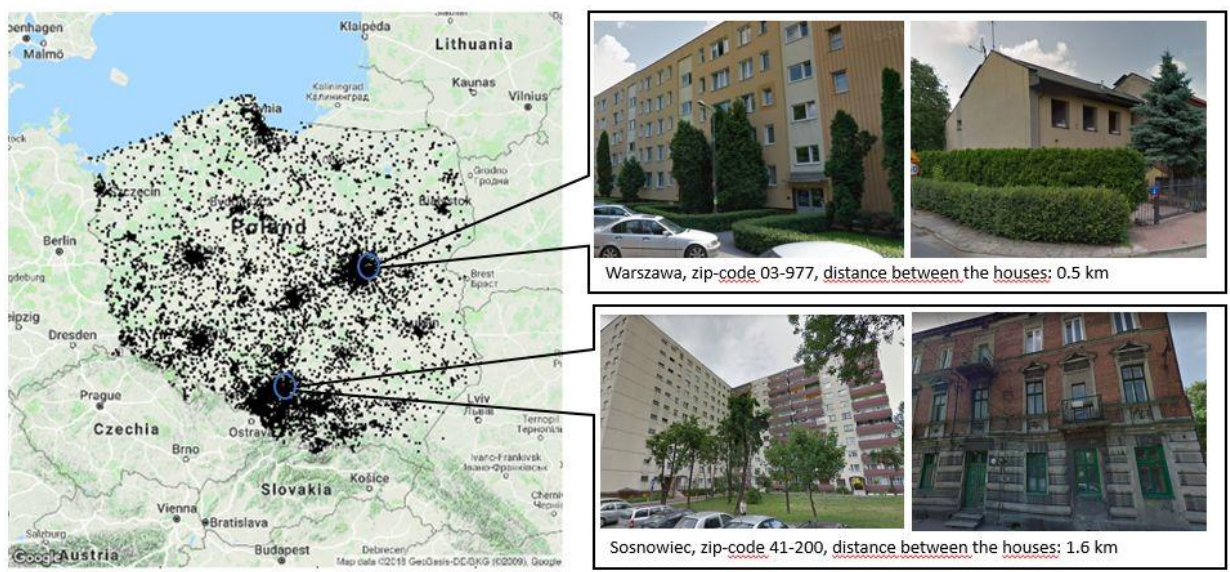

Figure 1. Examples of extremely different houses located in the same zip-code and residents of which have the same expected claim frequency by the current insurer's model.

Study of this insurance problem enabled following sociological and methodological discoveries: 1) features of the house correlate with the car accident risk of its resident,  2) compared to other uses of Google Street View for research, our variables are sourced from the address rather than aggregated by zip-code or district and they allow for new sociological discoveries at a very granular level, 3) variables extracted from the address (the image of a house) can be used in insurance and other industries, notably for price discrimination, 4) modern data collection and computational techniques, which allow for unprecedented exploitation of personal data, can outpace development of legislation and raise privacy threats.

**Results**

We examine a motor insurance dataset of 20,000 records—a random sample of an insurer's portfolio collected in Poland over the period January 2012 to December 2015. Each record represents characteristics of an insurance policy covering motor third party liability (MTPL) including the address of the policyholder, risk exposure defined as a fraction of the year in which the policy was active over the period 2013-2015 and the count of incurred property damage claims over the period 2013-2015. The insurer provided us also with the expected frequency of property damage claims for those policies, estimated by their current best-in-class risk model, that includes zoning based on the client's zip-code.

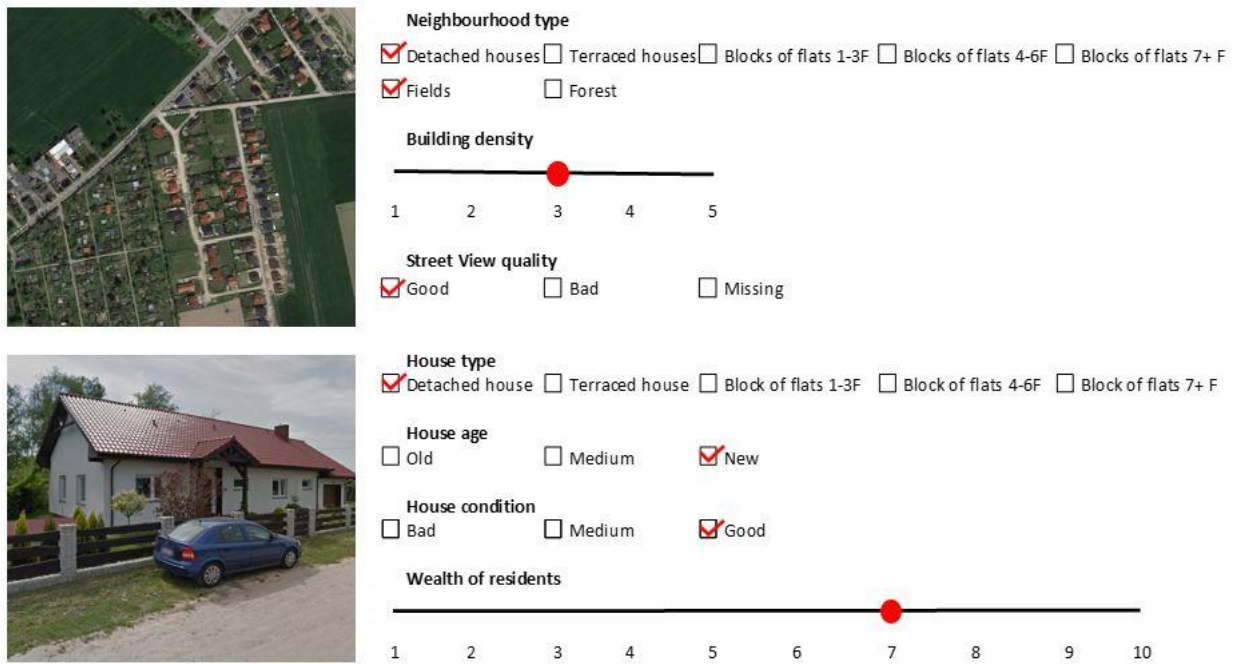

Figure 2. Features annotated from Google Satellite View and Google Street View image of a particular address.

We collect Google Satellite View and Google Street View images for the addresses provided in the database. Six experts annotated the following features of the houses visible in the images: their type, age, condition, estimated wealth of its residents, as well as type and density of other buildings from the neighborhood (Figure 2). Four out of six annotators gave moderately consistent answers for the common subsample of 500 addresses – Fleiss' kappa statistics indicate mostly moderate agreement among them (Table 1). These four annotators continued annotating remaining 19,371 addresses (we removed 129 addresses from the scope of this study as they were either foreign or could not be found by Google Maps), but this time each annotator was given a separate, randomly selected, set of addresses. We compared distributions of collected annotations and at the end applied small corrections to match the mean and standard deviation among all four annotators.

| variable | original granularity | inter-rater reliability | | risk model | |
|---|---|---|---|---|---|
| | | Fleiss' kappa | interpretation | granularity after simplification | p-value |
| Neighbourhood type | 7 types, multi-choice | 0.52 | moderate agreement | 2 | 0.01 |
| Building density | scale 1-5 | 0.50 | moderate agreement | not significant | |
| Street View quality | good / bad / missing | 0.79 | substantial agreement | 2 | 0.02 |
| House type | 5 types, single-choice | 0.69 | substantial agreement | 2 | 0.01 |
| House age | scale 1-3 | 0.51 | moderate agreement | 2 | 0.03 |
| House condition | scale 1-3 | 0.54 | moderate agreement | 2 | 0.04 |
| Wealth of residents | scale 1-10 | 0.32 | fair agreement | not significant | |

Table 1. Statistics for 7 newly created variables—original granularity, inter-rater reliability of 4 selected annotators on the common set of 500 observations and significance in our risk model after applying necessary simplifications.

Next, we estimated a Generalized Linear Model (GLM) to investigate the importance of newly created variables for risk prediction (*9*, *12*, *13*). We assume the following probabilistic model of claim frequency *f*, defined as the number of claims divided by risk exposure:

$$\log(\mathbb{E}(f)) = \log(\mathbb{E}(Y/exposure)) = \beta X$$

where $Y$ is a number of property damage claims within MTPL insurance following Poisson distribution, $X$ is a vector of independent variables and $\beta$ is the vector of coefficients.

For relative evaluation of the value added by our approach, we introduce three models:

- Model A (null model), where vector $X$ is $[1]$
- Model B (best-in-class insurer's model): where vector $X$ is $[1, X_1, , X_j]$
- Model C (our model): where vector $X$ is $[1, X_1, , X_j, X_{j+1}, , X_N]$

The insurer provided us with the realization of the model B for each record from the dataset. That model was estimated on a larger undisclosed dataset and contains *j* predictive variables (driver characteristics, vehicle characteristics, claim history, geographical zone, etc.). Using properties of GLMs we can decompose Model C into two parts: one corresponding to the Model B and one incorporating the new variables. We refer to the realization of the Model B multiplied by exposure as an offset (*14*) and do not estimate it. Therefore, Model C takes form

$$\log(\mathbb{E}(Y)) = \beta_0 + \beta_{j+1}X_{j+1} + ... + \beta_N X_N + \log(offset)$$

Intuitively, in this representation, the estimated coefficients $\beta_{j+1}, \ldots, \beta_N$ explain the signal that is not explained by the best-in-class risk model of the insurer (model B) and will also adjust for the earned exposure of the policy shorter than 1 year. We investigate if the values of these coefficients are non-zero, indicating that the variables we constructed provide additional predictive power to the model. We find that five out of seven newly created variables within this research are significant for predicting property damage MTPL claim frequency model, on top of many other rating variables used in the best-in-class insurer's model (Table 1).

We observe a significant variability of Gini coefficient for all A, B, C models – in particular for model A (null model with intercept only and no other variables selected) it varies from 20 to 38% within 20 resampling trials. We interpret it as the evidence that the dataset provided is extremely small (20,000 records) for modeling such rare events as property damage claims within MTPL insurance (average frequency of 5%).

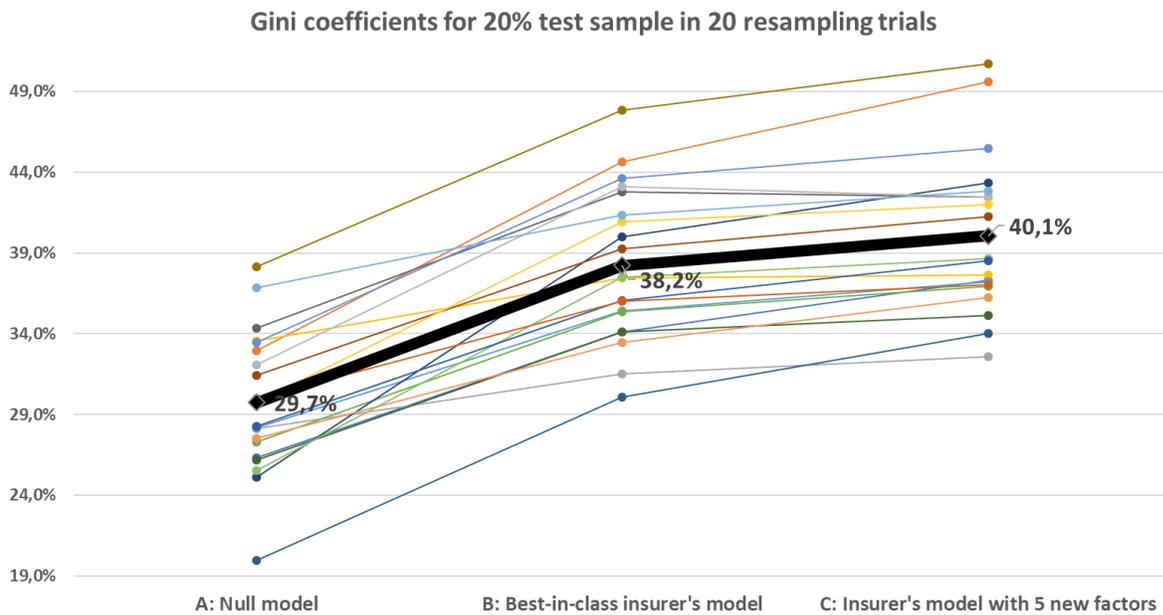

Figure 3. Gini coefficients obtained on 20% test sample in 20 bootstrapping trials from the null model (A), the best-in-class insurer's model (B) and our model with newly created variables (C).

Despite the high volatility of data, adding our five simple variables to the insurer's model improves its performance in 18 out of 20 resampling trials and the average improvement of the Gini coefficient is nearly 2 percentage points (from 38.2% to 40.1%). To put this value into perspective, the best-in-class insurer's model fitted on much bigger dataset and including a broad selection of variables (e.g. driver characteristics, car characteristics, claim history and geographical zones based on the client's zip-code) improves the Gini coefficient versus null model by 8 percentage points from ~30% to ~38% (Figure 3).

**Discussion**

We found that features visible on a picture of a house can be predictive of car accident risk, independently from classically used variables such as age, or zip code. This finding is not only a step towards more granular risk prediction models, but also illustrates a novel approach to social science, where the real-world granular data is collected and analyzed at scale.

From the practical perspective of insurance companies, the results we present are remarkably powerful, when compared to the best-in-class insurance model. Our 5 variables, containing already some bias from the imperfect annotation, improve Gini coefficient by nearly 2 percentage points, which is massive, comparing to the improvement of 8 percentage points brought by numerous variables that the insurer has already been using in his best-in-class risk model. The insurance industry could be quickly followed by the banks, as there is a proven correlation between insurance risk models and credit risk scoring (*15*). The approach itself to extract valuable information from Google Street View opens a variety of opportunities not only for the financial sector. Any company that collects clients' addresses could adopt our methodology and deep learning technology enables to make it in an automated way on a massive scale (*16*).

Such practice, however, raises concerns about the privacy of data stored in publicly available Google Street View, Microsoft Bing Streetside, Mapillary, or equivalent privately-held datasets like CycloMedia. The consent given by the clients to the company to store their addresses does not necessarily mean a consent to store information about the appearance of their houses. In particular, features of the house may be a proxy of ethnicity, religion or other characteristics associated with a social status of a person (*17*, *18*) which are forbidden by the law to be used for any discrimination, e.g. price discrimination in certain jurisdictions (*19*). Fast development of modern data collection and computational techniques allows for unprecedented exploitation of various data of clients being not even aware of it (*20*) and development of corresponding legislation in this matter seems to be outpaced.

The methods we present could be substantially improved by employing more annotators for the same set of the images. Potentially the average or ensemble of their answers would match the reality better than an annotation of a single person (*21*, *22*). Our model performance could be different on the data of another insurer or in another country, however, to the best of our knowledge, most insurance companies do aggregate risk at the zip-code or regional level, so our variables are still likely to provide an additional signal. Another limitation is the small size of the dataset provided by the insurance company, but we reduced this problem using bootstrapping and by using elementary modelling techniques such as the generalized linear models.

**References**

1. M. Finer *et al.*, Combating deforestation: From satellite to intervention. *Science*. **360**, 1303–1305 (2018).

2. N. Jean *et al.*, Combining satellite imagery and machine learning to predict poverty. *Science*. **353**, 790–794 (2016).

3. A. Esteva *et al.*, Dermatologist-level classification of skin cancer with deep neural networks. *Nature*. **542**, 115–118 (2017).

4. P. Lakhani, B. Sundaram, Deep Learning at Chest Radiography: Automated Classification of Pulmonary Tuberculosis by Using Convolutional Neural Networks. *Radiology*. **284**, 574–582 (2017).

5. D. Anguelov *et al.*, Google Street View: Capturing the World at Street Level. *Computer* . **43**, 32–38 (2010).

6. T. Gebru *et al.*, Using deep learning and Google Street View to estimate the demographic makeup of neighborhoods across the United States. *Proc. Natl. Acad. Sci. U. S. A.* **114**, 13108–13113 (2017).

7. R. Goel *et al.*, Estimating city-level travel patterns using street imagery: A case study of using Google Street View in Britain. *PLoS One*. **13**, e0196521 (2018).

8. V. O. Andersson, M. A. F. Birck, R. M. Araujo, in *Computational Neuroscience* (Springer International Publishing, 2017), pp. 81–93.

9. G. Werner, C. Modlin, in *Casualty Actuarial Society* (2010).

10. G. Taylor, Geographic Premium Rating by Whittaker Spatial Smoothing. *ASTIN Bulletin: The Journal of the IAA*. **31**, 147–160 (2001).

11. J. Yao, Clustering in Ratemaking: Applications in Territories Clustering. *Casualty Actuarial Society Discussion Paper Program Casualty Actuarial Society-Arlington, Virginia*, 170–192 (2008).

12. G. A. Spedicato, C. Dutang, L. Petrini, Machine Learning Methods to Perform Pricing Optimization. A Comparison with Standard GLMs. *Variance: Advancing the Science of Risk*. **111** (2018).

13. I. Kolyshkina, S. Wong, S. Lim, in *Casualty Actuarial Society* (2004), pp. 279–290.

14. J. Yan, J. Guszcza, M. Flynn, C.-S. P. Wu, in *Casualty Actuarial Society E-Forum, Winter 2009* (2009), p. 366.

15. L. L. Golden, P. L. Brockett, J. Ai, B. Kellison, Empirical Evidence on the Use of Credit Scoring for Predicting Insurance Losses with Psycho-social and Biochemical Explanations. *N. Am. Actuar. J.* **20**, 233–251 (2016).

16. B. Zhou, A. Lapedriza, J. Xiao, A. Torralba, A. Oliva, in *Advances in Neural Information Processing Systems 27*, Z. Ghahramani, M. Welling, C. Cortes, N. D. Lawrence, K. Q. Weinberger, Eds. (Curran Associates, Inc., 2014), pp. 487–495.

17. A. R. Gillis, Population Density and Social Pathology: The Case of Building Type, Social Allowance and Juvenile Delinquency. *Soc. Forces*. **53**, 306–314 (1974).

18. E. R. Braver, Race, Hispanic origin, and socioeconomic status in relation to motor vehicle occupant death rates and risk factors among adults. *Accid. Anal. Prev.* **35**, 295–309 (2003).

19. J. Gaulding, Race Sex and Genetic Discrimination in Insurance: What's Fair. *Cornell Law Rev.* **80**, 1646 (1994).

20. M. J. Blitz, The right to map (and avoid being mapped): reconceiving First Amendment protection for information-gathering in the age of Google Earth. *Columbia Sci. Technol. Law Rev.* **14**, 115 (2012).

21. L. Tran-Thanh, S. Stein, A. Rogers, N. R. Jennings, Efficient crowdsourcing of unknown experts using bounded multi-armed bandits. *Artif. Intell.* **214**, 89–111 (2014).

22. R. M. Levenson, E. A. Krupinski, V. M. Navarro, E. A. Wasserman, Pigeons (Columba livia) as Trainable Observers of Pathology and Radiology Breast Cancer Images. *PLoS One*. **10**, e0141357 (2015).

23. M. Goldburd, A. Khare, C. D. Tevet, in *Casualty Actuarial Society* (2016).

24. P. Cizek, W. K. Härdle, R. Weron, *Statistical Tools for Finance and Insurance* (Springer Science & Business Media, 2005).

25. E. W. Frees, G. Meyers, A. D. Cummings, Summarizing Insurance Scores Using a Gini Index. *J. Am. Stat. Assoc.* **106**, 1085–1098 (2011).

26. M. O. Lorenz, Methods of Measuring the Concentration of Wealth. *Publications of the American Statistical Association*. **9**, 209–219 (1905).

27. C. Gini, Measurement of Inequality of Incomes. *Econ. J. Nepal*. **31**, 124–126 (1921).

**Funding:** Łukasz Kidziński is partly sponsored by the Mobilize Center at Stanford, a National Institutes of Health Big Data to Knowledge (BD2K) Center of Excellence supported through Grant U54EB020405

**Author contributions:** Authors contributed equally.

**Competing interest:** Authors declare no competing interests.

**Data and materials availability:** All data required to understand and assess the conclusions of this research are available in the main text and supplementary materials. Insurance claims data used for the study is privately owned and contains sensitive client observations precluding publication.